\documentclass[10pt,conference]{IEEEtran}
\IEEEoverridecommandlockouts
\usepackage{cite}
\usepackage{amssymb,amsfonts}
\usepackage{textcomp}
\usepackage[dvipsnames]{xcolor}
\usepackage{hyperref}
\usepackage{setspace}
\usepackage{graphicx}
\usepackage[spanish,ngerman,french,main=english]{babel}
\usepackage[utf8]{inputenc}
\usepackage[autostyle,english=american]{csquotes}
\usepackage{caption}
\MakeOuterQuote{"}
\usepackage{amsmath}

\usepackage{algorithmic}
\usepackage{array}
\usepackage{soul}
\usepackage{multirow}
\usepackage{xtab,booktabs,rotating}
\usepackage{enumitem}
\usepackage[caption=true,font=footnotesize]{subfig}
\usepackage{xhfill}
\usepackage{url}
\usepackage{todonotes}
\usepackage[markup=underlined]{changes}
\definechangesauthor[color=NavyBlue]{YD}
\setcommentmarkup{\todo[color={authorcolor!20},size=\scriptsize]{#3: #1}}
\usepackage{microtype}        % Improved Tracking and Kerning

\def\BibTeX{{\rm B\kern-.05em{\sc i\kern-.025em b}\kern-.08em
    T\kern-.1667em\lower.7ex\hbox{E}\kern-.125emX}}

\makeatletter
\def\@IEEEpubidpullup{8\baselineskip}
\makeatother

\graphicspath{{Images/}}

\begin{document}

% \IEEEoverridecommandlockouts
% \IEEEpubid{
% \parbox{\columnwidth}{\vspace{-4\baselineskip}Permission to make digital or hard copies of all or
% part of this work for personal or classroom use is granted without fee provided that copies are not
% made or distributed for profit or commercial advantage and that copies bear this notice and the full
% citation on the first page. Copyrights for components of this work owned by others than ACM must
% be honored. Abstracting with credit is permitted. To copy otherwise, or republish, to post on servers
% or to redistribute to lists, requires prior specific permission and/or a fee. Request permissions from
% \href{mailto:permissions@acm.org}{permissions@acm.org}.\hfill\vspace{-0.8\baselineskip}\\
% \begin{spacing}{1.2}
% \small\textit{ASONAM '21}, November 8-11, 2021, Virtual Event, Netherlands \\
% \copyright\space2021 Association for Computing Machinery. \\
% ACM ISBN 978-1-4503-9128-3/21/11?/\$15.00 \\
% \url{http://dx.doi.org/10.1145/3487351.3488353}
% \end{spacing}
% \hfill}
% \hspace{0.9\columnsep}\makebox[\columnwidth]{\hfill}}
% \IEEEpubidadjcol

\title{Interpretable Business Survival Prediction}

\author{
\IEEEauthorblockN{
Anish K. Vallapuram\IEEEauthorrefmark{1}, Nikhil Nanda\IEEEauthorrefmark{1}, Young D. Kwon\IEEEauthorrefmark{2}, and Pan Hui\IEEEauthorrefmark{1}\IEEEauthorrefmark{3}
}
\IEEEauthorblockA{\IEEEauthorrefmark{1}Hong Kong University of Science and Technology, Hong Kong SAR}
\IEEEauthorblockA{\IEEEauthorrefmark{2}University of Cambridge, United Kingdom}
\IEEEauthorblockA{\IEEEauthorrefmark{3}University of Helsinki, Finland}
Email: \{akvallapuram, nnanda\}@connect.ust.hk \quad ydk21@cam.ac.kr \quad panhui@ust.hk
}

\maketitle

\begin{abstract}

The survival of a business is undeniably pertinent to its success. A key factor contributing to its continuity depends on its customers. The surge of location-based social networks such as Yelp, Dianping, and Foursquare has paved the way for leveraging user-generated content on these platforms to predict business survival. Prior works in this area have developed several quantitative features to capture geography and user mobility among businesses. However, the development of qualitative features is minimal. In this work, we thus perform extensive feature engineering across four feature sets, namely, geography, user mobility, business attributes, and linguistic modelling to develop classifiers for business survival prediction. We additionally employ an interpretability framework to generate explanations and qualitatively assess the classifiers' predictions. Experimentation among the feature sets reveals that qualitative features including business attributes and linguistic features have the highest predictive power, achieving AUC scores of 0.72 and 0.67, respectively. Furthermore, the explanations generated by the interpretability framework demonstrate that these models can potentially identify the reasons from review texts for the survival of a business.

\end{abstract}

% \IEEEpeerreviewmaketitle

% \begin{IEEEkeywords}
% component, formatting, style, styling, insert
% \end{IEEEkeywords}

% 1. intro
% 2. related work
% 3. Geographical modelling
%   3.1 features
%   3.2 results
% 4. Linguistic modelling
%   4.1 features
%   4.2 results
% 5. Visual modelling
%   5.1 features
%   5.2 results
% 6. Combined Models and Interpretability
%   6.1 results of all features combined
%   6.2 interpretability
% 7. Conclusions and future work

%%%%%%%%%%%%%%%%%%%%%%%%%%%%%%%%%%%%%%%%%%
% Introduction
%%%%%%%%%%%%%%%%%%%%%%%%%%%%%%%%%%%%%%%%%%
\section{Introduction}\label{sec:introduction}

``Customer Is King'' is a ubiquitous motto followed by several businesses. With the rise of technology, online reviews regarding businesses have become increasingly relevant as a measurement of customer satisfaction. A working paper from Harvard Business School \cite{hbs_yelp} concluded that non-chain restaurants in Yelp gain 5\% to 9\% in revenue with a one-star increase in rating. The online perception of businesses on social media thus has a bearing on their survival.

Recently, several works have leveraged location-based social network (LBSN) platforms to engineer features and inference models to predict business survival. Tayeen et al. \cite{location3} develop features to capture housing affordability, education, and tourism near restaurants in Yelp to predict their success. D'Silva et al. \cite{dsilva} and Lian et al. \cite{hetero_info} model the geography and customer visitation patterns among businesses in Foursquare and Dianping, respectively. However, All the aforementioned features are quantitative involving complex formulations. Developing qualitative features are highly sought after as they are more comprehensible. Still, investigations on extracting qualitative features are limited to Lian et al. \cite{hetero_info} who reveal that reviews possess high predictive power in this prediction task. In this work, we thus engineer an extensive feature set including quantitative features from geography and user mobility and qualitative features from (i) \textbf{business attributes} and (ii) \textbf{linguistic modelling}.

Nevertheless, we believe that simply predicting the survival of a business is not adequate. The classifiers must be augmented with interpretability to generate explanations for the predictions. These explanations can provide avenues for businesses to improve upon. For example, if the classifier predicts a lower chance of a business's survival due to its pricing, the business may adjust to a more competitive pricing. Furthermore, applying interpretability to the classifier allows us to qualitatively assess whether it identifies human-intuitive features about business survival \cite{clever_hans, molnar2020interpretable}.

Thus, this work tackles the following research questions:
\begin{itemize}
    \item \textbf{RQ1:} What qualitative features can be effectively incorporated to the business survival prediction problem?
    \item \textbf{RQ2:} Can an interpretability framework provide human-intuitive explanations to the predictions of the classifiers?
\end{itemize}

\textbf{First of all}, we attempt to perform business survival prediction from the Yelp dataset. While the dataset includes businesses from several categories, we only consider restaurants. This work can still be generalised to other business categories. We then formulate a survival prediction task. We only consider restaurants in the dataset that are open till the end of 2017 and predict their survival by the end of 2019. This yields a total of 36,190 restaurants with 12.8\% of them labelled dead by the end of 2019. 

\textbf{Then,} we follow feature engineering performed by D'Silva et al. \cite{dsilva} to capture the geography and user mobility of the restaurants. The geographical features primarily represent competition, business category counts and attractiveness of the neighbourhood of a restaurant. The user mobility features represent the transitions of customers among these businesses by modeling the inflow and outflow, temporal popularity skew, temporal alignment with competitors and visit trends for each restaurant. We then employ several machine learning models to predict business survival. The best-performing model yields an AUC score of 0.62 with reasonable explanations for the predictions. We adopt the representative explainable machine learning method, LIME~\cite{LIME}, which is a model-agnostic framework to generate explanations for predictions made by our models.

\textbf{Next,} our work explores several business attributes provided in the Yelp dataset for this task. These attributes include \textit{Pricing}, \textit{Ambience}, \textit{Restrictions}, \textit{Amenities}, \textit{Services} as well as \textit{Image and Review counts}. Improved performance is achieved when compared with survival prediction using geography and user mobility features. The best model achieved an AUC score of 0.72 along with qualitative interpretations of the predictions.

\textbf{Additionally}, for linguistic modelling, we follow Lian et al. \cite{hetero_info} and take a bag-of-words approach for predicting business survival through customer reviews. As a further improvement, we also implement a recurrent neural network to better capture the long-term dependencies within the language of the reviews. We also explore an alternative prediction task of sentiment analysis of customers from reviews to better capture their perceptions regarding restaurants. Empirical results suggest that the sentiment analysis task yields a better AUC score of 0.98 while survival prediction only yields an AUC score of 0.67. Explanations from the linguistic models also demonstrate that the sentiment analysis models effectively capture customers' impressions than the survival prediction counterparts.

% Ablation Study results
\textbf{Finally}, we perform an ablation study to evaluate the performance of combining different features. Results reveal that business attributes and linguistic features perform the best among all the different combinations of the developed predictors, demonstrating the effectiveness of qualitative feature sets. Yet, these two features perform significantly better individually.

% \young{the paragraph below is about paper organization. we can always remove this part if we need some space.}
The rest of this paper is organised as follows. Section \ref{sec:Related Work} discusses some recent literature that inspired the features and models employed in this work. Section \ref{sec:Data and Problem Statement} details the dataset utilised and formulates our problem statement. Sections \ref{sec:Geo_and_User} through \ref{sec:Linguistic} entail the feature extraction and prediction models implemented across the four feature sets and analyses of the results. We conclude and call for future work in Section \ref{sec:Conclusion}. 

%%%%%%%%%%%%%%%%%%%%%%%%%%%%%%%%%%%%%%%%%%
% Related Work
%%%%%%%%%%%%%%%%%%%%%%%%%%%%%%%%%%%%%%%%%%
\section{Related Work}\label{sec:Related Work}

Early literature in this area originates from the finance and management field that investigates bankruptcy prediction \cite{olsen, kim_gu, upneja}. However, these works utilise several financial variables which are difficult to acquire. 

Alternatively, several recent investigations have leveraged the data generated by LBSNs for business survival prediction \cite{on_the_brink, olympics, dsilva, location3, hetero_info}. Works that are closest in spirit to our problem context include D'Silva et al. \cite{dsilva} and Lian et al. \cite{hetero_info} who model the geography and customer visitation patterns among businesses in Foursquare and Dianping respectively. However, their proposed features are quantitative involving complex formulations. Developing more qualitative features are highly sought after as they are more comprehensible. Still, investigations on extracting qualitative features are limited to Lian et al. \cite{hetero_info} who reveal that reviews possess high predictive power in this prediction task which bolsters our argument on the inclusion of qualitative features from reviews for this prediction task. 

Another highly relevant area for this work is interpretable machine learning. The literature of interpretability is primarily focused on two approaches to provide explanations or more information about the models' predictions~\cite{molnar2020interpretable}. One direction is to adopt "weight-box" models that are inherently explainable such as decision tree, logistic regression models. The other is to develop methods such as PDP~\cite{PDP}, LIME~\cite{LIME}, SHAP~\cite{SHAP} to explain "black-box" models that produce a mere prediction result (e.g., deep neural networks).
LIME is a model-agnostic framework that approximates the local neighbourhood of a prediction of interest to generate explanations for this prediction from a model. 
In this work, we employ LIME which is a representative framework for interpretable machine learning~\cite{Casual_Inter} to obtain interpretability for various feature sets and different models that we have employed.

% [paper objectives/contributions]
At a glance, this work aims to develop interpretable classifiers to predict business survival using insights cultivated from user-generated content on an LBSN and derive the explanations for the classifiers' predictions. 

%%%%%%%%%%%%%%%%%%%%%%%%%%%%%%%%%%%
% Data and Problem Statement
%%%%%%%%%%%%%%%%%%%%%%%%%%%%%%%%%%%
\section{Data and Problem Statement}\label{sec:Data and Problem Statement}

In this section, we first describe the datasets employed for our analysis including some preliminary statistics on its attributes. We then formalise our business survival prediction task based on these datasets.

%----------------------------------
% Yelp Dataset
%----------------------------------
\subsection{Yelp Dataset}

The principal dataset utilised in this work is the publicly available Yelp dataset. Yelp is the world’s largest platform for customer reviews and covers a wide range of data including business, geographical and meta information and customer engagement through reviews, check-ins and image uploads. Table~\ref{table:basic_stats} provides some basic statistics on this dataset.

% Table
\begin{table}[h!]
\centering
\begin{tabular}{cccc}
\toprule
 \# businesses & \# reviews & \# check-ins & \# images \\
\midrule
 209,393 & 7,734,455 & 21,003,777 & 200,000 \\
\bottomrule
\end{tabular}
\caption{Basic statistics regarding the Yelp dataset 2019 version.}
\label{table:basic_stats}
\end{table}

\begin{figure}[t!]
    \centering
    \includegraphics[scale=0.55]{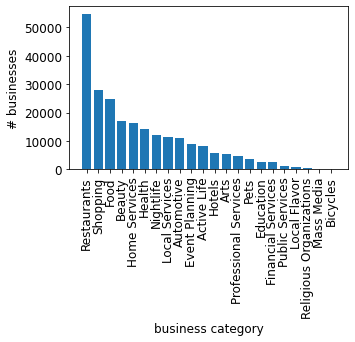}
    \caption{Number of businesses in each category in Yelp.}
    \label{fig:cate_counts}
\end{figure}

\begin{figure}[t!]
    \centering
    \includegraphics[scale=0.5]{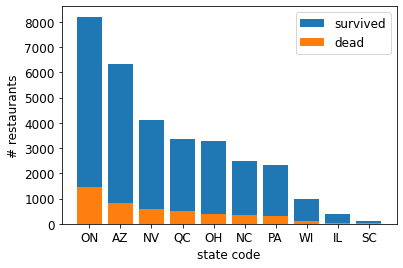}
    \caption{Number of survived and dead businesses by states. }
    \label{fig:state_count}
\end{figure}

The businesses in the Yelp dataset are further divided into 22 categories as summarised in Figure~\ref{fig:cate_counts}. Furthermore, the dataset has a diverse set of attributes catered to the specific category of business. For example, restaurants have attributes such as \textit{TakeOut, Attire, Delivery}, while hair salons have attributes such as \textit{HairSpecializesIn}. There is no intersection in features among these two categories of businesses, i.e. \textit{HairSpecializesIn} is set to null for restaurants and vice versa. This could lead to very sparse data. Hence, this study is only focused on the restaurants' category because this category not only has the largest number of businesses but also the largest number of attributes, which then would aid our prediction in a way that it gives more data to process. Finally, the restaurants are also geographically grouped into 9 states and provinces across the United States and Canada as shown in Figure~\ref{fig:state_count}. This allows us to consider businesses under various socioeconomic conditions.

% figure

% figure

%----------------------------------
% Survival Prediction
%----------------------------------
\subsection{Survival Prediction} 

\begin{figure}[t]
    \centering
    \includegraphics[scale=0.19]{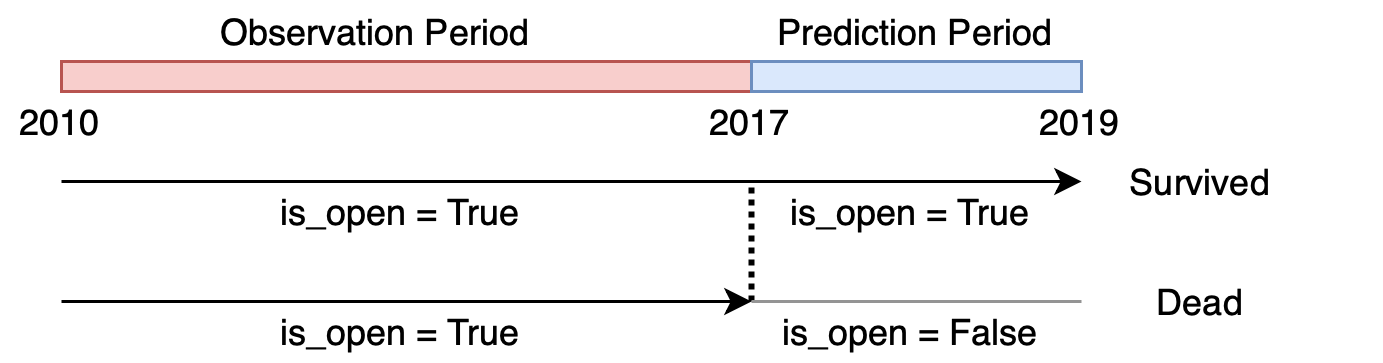}
    \caption{Observation and prediction period for Yelp dataset.}
    \label{fig:survival_label}
\end{figure}

% Expand on past works if necessary.  
We now turn our attention to the definition of the business survival prediction problem. The objective is to predict whether a business will remain open or face closure after a fixed duration of time given several attributes regarding the business. The Yelp dataset provides a boolean attribute \textit{is\_open} which indicates the closure status of a business indefinitely in time. The date of the business closure must be known to define the start-of-future (SOF) for the prediction task. Hence, we leverage the fact that Yelp updates its dataset annually. This study employs two versions of the Yelp dataset from 2017 and 2019 and the survival label is determined as follows. We define the observation period to be from the date of inception of Yelp to the end of 2017 and the prediction period from the start of 2018 till the end of 2019. We only consider businesses that have \textit{is\_open} set to \textit{True} in the observation period. A business is labelled as survived if it is then open at the end of the prediction period and dead otherwise. Figure~\ref{fig:survival_label} is a visual depiction of our problem statement. Based on the above timeframe, we obtain a dataset with the composition of survived and dead restaurants as shown in Table~\ref{table:num_restaurants}. 

% Figure

% Table
\begin{table}[t]
\centering
\begin{tabular}{lrr}
\toprule
 Dataset & \# open & \# closed \\
\midrule
 Yelp (2017) & 36,190 & 13,106 \\
 Yelp (2019) & 31,617 & 17,679 \\
 \hline
  & \# survived & \# dead \\
 \hline
 Ours & 31,546 & 4,644 \\
\bottomrule
\end{tabular}
\caption{Number of open and closed restaurants from different versions of the Yelp dataset and the number of survived and dead restaurant in our dataset.}
\label{table:num_restaurants}
\end{table}

%%%%%%%%%%%%%%%%%%%%%%%%%%%%%%%%%%%
% Geography and User Mobility modelling
%%%%%%%%%%%%%%%%%%%%%%%%%%%%%%%%%%%
\section{Geography and User Mobility Modelling}\label{sec:Geo_and_User}
As investigated by \cite{dsilva, hetero_info}, the performance of a restaurant is highly dependent on its location and customer satisfaction. The two works thus perform extensive feature engineering to model these two aspects of a business's success. Following D'Silva et al. \cite{dsilva}, we develop locality profiles and user mobility features for restaurants in the Yelp dataset. 

%----------------------------------
% Locality Profiles
%----------------------------------
\subsection{Locality Profiles}
The locality or the neighbourhood of a restaurant $r$ is defined as the set of other businesses $V_{r}$ within a radius of 500m based on prior work \cite{500m}. Though the prediction is performed only for restaurants, businesses of all categories are considered for the neighbourhood. The following features are then employed.  

\subsubsection{Competition} The competition to a restaurant is the ratio of businesses in the neighbourhood that are restaurants. 

\begin{equation}
    f_{r}^{C} = \frac{|\{v_{i}: \mathcal{R}(v_{i}) \cap v_{i} \in V_{r}\}|}{|V_{r}|}
\end{equation}

$\mathcal{R}(u)$ returns true if $u$ is a restaurant. 

\subsubsection{Specific Competition} 
\begin{figure}
    \centering
    \includegraphics[scale=0.52]{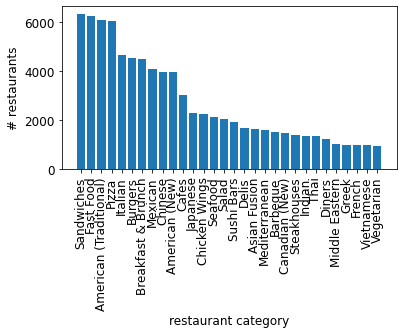}
    \caption{Top 30 cuisines by the number of restaurants.}
    \label{fig:sub_cate_counts}
\end{figure}

The restaurants' category in the Yelp dataset is further divided into 145 subcategories based on the cuisines served and Figure~\ref{fig:sub_cate_counts} depicts the number of restaurants serving the top 30 cuisines in the dataset. The specific competition for a business is then the ratio of restaurants in its neighbourhood serving the same cuisine. 

\begin{equation}
    f_{r}^{SC} = \frac{|\{v_{i}: \mathcal{SC}(v_{i}, v_{r}) \cap v_{i} \in V_{r}\}|}{|\{v_{i}: \mathcal{R}(v_{i}) \cap v_{i} \in V_{r}\}|}
\end{equation}

$\mathcal{SC}(u, v)$ is true if restaurants $u$ and $v$ serve the same cuisine. 

\subsubsection{Category Counts} The business category and the restaurant sub-category counts within the neighbourhoods are also tallied as features. The category $C_{r}$ and sub-category $SC_{r}$ counts are represented as 22-element and 145-element vectors respectively.

% Figure

\subsubsection{Place Entropy} To represent the diversity in competition within the neighbourhood, the place entropy %is measured which is 
measures
the Shannon Entropy of the probabilities of the business categories in the restaurant's neighbourhood. 

\begin{equation}
    f_{r}^{PE} = - \sum_{c=1}^{22} p_{c} * \ln(p_{c})
\end{equation}

\subsubsection{Neighbourhood Attractiveness} This metric is developed to capture a more global view of a restaurant's neighbourhood compared to other neighbourhoods by utilising a TF-IDF approach. This metric is measured both at the business category and restaurant sub-category level. The documents are analogous to the business categories and sub-categories and their constituent term frequencies are analogous to the category $C_{r}$ and sub-category $SC_{r}$ counts discussed earlier. Thus, this metric yields two vectors at the category and sub-category levels with the same dimensions. 

\begin{eqnarray}
    f_{r}^{NAC} &=& TFIDF(C_{r}) \\ 
    f_{r}^{NASC} &=& TFIDF(SC_{r})
\end{eqnarray}

%----------------------------------
% User Mobility
%----------------------------------
\subsection{User Mobility} 
The customer visit patterns among these restaurants will now be modelled. Prior works \cite{dsilva, hetero_info} employ datasets from the platforms Foursquare and Dianping respectively which are similar to Yelp. These platforms include an additional type of customer engagement called transitions. A transition occurs when a user makes consecutive check-ins to the two businesses within a day. The check-ins in the Yelp dataset cannot be identified by the users that made them, yet the reviews can be identified by users. Hence, we utilise all consecutive pairs of reviews from a user as transitions from the user. 

% LimeTabularExplainer
\begin{figure*}[ht!]
 \centering
 \begin{minipage}[b]{0.45\linewidth}
 \includegraphics[scale=0.4]{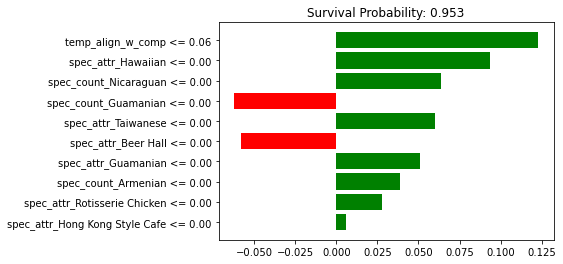}
  \caption*{(a) Central Diner and Grille, PA}
 \end{minipage}
 \quad
 \begin{minipage}[b]{0.45\linewidth}
    \includegraphics[scale=0.4]{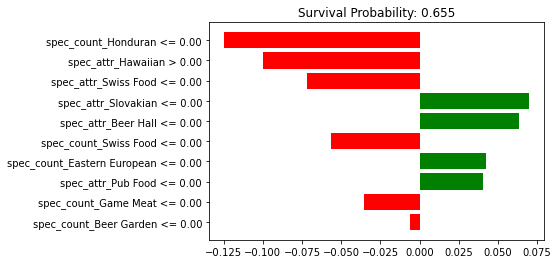}
   \caption*{(b) Planet Smoothie, AZ}
\end{minipage}
\caption{Top 10 most important geographical and user mobility features for survival predictions of two restaurants. Important features for survival are coloured green and death are coloured red.}
\label{fig:geo_user_exp}
\end{figure*}

\subsubsection{Inflow and Outflow} This class of metrics measure the number of transitions to and from a restaurant during the restaurant's lifespan. Since the start date of a business is not available in Yelp, it is determined from the time of the first check-in or review received by the restaurant. The inflow and outflow are then formally defined as 

\begin{eqnarray}
    f_{r}^{IN} &=& \frac{1}{M} \sum_{j=0}^{|V_{r}|} \mathcal{T}(v_j, r) \\
f_{r}^{OUT} &=& \frac{1}{M} \sum_{j=0}^{|V_{r}|} \mathcal{T}(v_r, r)
\end{eqnarray}

where M is the lifespan of the restaurant in months and $\mathcal{T}(u, v)$ is an identity function which returns 1 if there exists a transition between businesses $u$ and $v$ and 0 otherwise.  

\subsubsection{Distance and Speed Travel} The average distance and speed travelled across all transitions occurring to and from a restaurant are also measured. 

\begin{eqnarray}
    f_{r}^{D_{to}} &=& \frac{\sum_{v_j \in \mathcal{T}_{r}^{to}} \mathcal{D}(v_j, r)}{|\mathcal{T}_{r}^{to}|} \\
    f_{r}^{D_{from}} &=& \frac{\sum_{v_j \in \mathcal{T}_{r}^{from}} \mathcal{D}(r, v_j)}{|\mathcal{T}_{r}^{from}|} \\
    f_{r}^{S_{to}} &=& \frac{\sum_{v_j \in \mathcal{T}_{r}^{to}} \mathcal{S}(v_j, r)}{|\mathcal{T}_{r}^{to}|} \\
    f_{r}^{S_{from}} &=& \frac{\sum_{v_j \in \mathcal{T}_{r}^{from}} \mathcal{S}(r, v_j)}{|\mathcal{T}_{r}^{from}|}
\end{eqnarray}

where $\mathcal{T}_{r}^{to}$ and $\mathcal{T}_{r}^{from}$ are sets of businesses with transitions to and from the restaurant and $\mathcal{D}$ and $\mathcal{S}$ return the distance and speed of transitions.

\subsubsection{Temporal Popularity Skew} The temporal profile of a business $h$ is defined as the normalised number of check-ins for each hour of the day and is represented as a 24-element vector. The temporal popularity skew then measures the entropy of the restaurant's temporal profile to determine if check-ins to the restaurant are local to a few hours within the day or spread evenly across the day. 

\begin{equation}
    f_{r}^{PS} = - \sum_{i=1}^{24} h_{i} * \ln(h_{i})
\end{equation}

\subsubsection{Temporal Alignment with Competitors} This metric measures the distance between the temporal profile of the restaurant $h$ and the aggregated temporal profile of the businesses within the neighbourhood $H$. The aggregation of temporal profiles is the average number of check-ins for each hour of the day across the neighbouring businesses. 

\begin{equation}
    f_{r}^{AC} = \sum_{i=1}^{24} (h_{i} - H_{i})^2
\end{equation}

\subsubsection{Visit Trend} The visit trend determines the trend of the number of check-ins per month for a restaurant during the last six months of the observation period. This metric is measured by determining the slope of the linear trend between the month and the number of check-ins. 

\begin{equation}
    f_{r}^{VT} = \frac{C_{t}(r) - b}{t}
\end{equation}

where $C_{t}(r)$ returns the number of check-ins received by restaurant $r$ over the time-span $t$. 

%----------------------------------
% Prediction Model
%----------------------------------
\subsection{Prediction Models}
\begin{figure}[t!]
    \centering
    \includegraphics[scale=0.5]{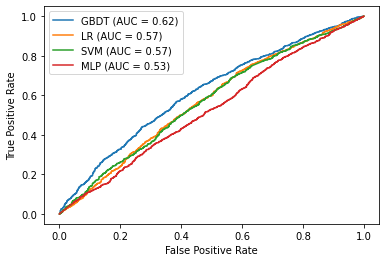}
    \caption{ROC curves for geography and user mobility predictors. Positive class refers to survived label.}
    \label{fig:roc_curve}
\end{figure}

Since this is an imbalanced classification task as evidenced by Table~\ref{table:num_restaurants}, we employ the SMOTE algorithm \cite{SMOTE} to synthetically oversample the dead restaurant instances and evaluate the prediction models using the Area Under the ROC Curve (AUC) \cite{auc_roc}. Both works \cite{dsilva, hetero_info} employ four models namely Gradient Boosting Decision Trees (GBDT), Logistic Regression (LR), Support Vector Machine (SVM) and Multilayer Perceptron (MLP). Among them, GBDT performs best as demonstrated in Figure~\ref{fig:roc_curve} with a reasonable AUC score of 0.62. 

%----------------------------------
% Explaining Results
%----------------------------------
\subsection{Explaining Results}
The focus of this work is not limited to demonstrating prediction models that outperform baselines. Qualitative explanations for the predictions must also be generated which would be highly insightful to business owners. To this end, the \textit{LimeTabularExplainer} is employed on the GBDT model and predictions of certain example restaurants were explored. In most cases, it was observed that the neighbourhood attractiveness to different business categories and restaurant sub-categories constituted the top 10 most important features for the survival predictions. 

Two examples were particularly interesting. As shown in Figure~\ref{fig:geo_user_exp}, a location of \textit{Planet Smoothie} (the third largest American smoothie chain) in Arizona was predicted to survive with very low confidence of 65.5 \% due to the presence of specific cuisines or the lack thereof in its neighbourhood. A more intriguing example is a seafood diner called \textit{Central Diner} in Pennsylvania was predicted with very high confidence to survive due to its temporal alignment with its competitors and the lack of specific cuisines in its neighbourhood.

%%%%%%%%%%%%%%%%%%%%%%%%%%%%%%%%%%%
% Business Attributes
%%%%%%%%%%%%%%%%%%%%%%%%%%%%%%%%%%%
\section{Business Attributes}\label{sec:Attributes}
As discussed in section~\ref{sec:Data and Problem Statement}, the Yelp dataset provides a plethora of business attributes. This work proposes to leverage these business attributes to predict business survival. 

% LimeTabularExplainer
\begin{figure*}[t!]
 \centering
 \begin{minipage}[b]{0.45\linewidth}
 \includegraphics[scale=0.4]{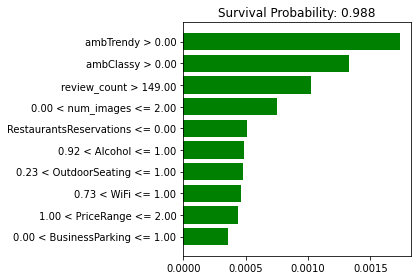}
  \caption*{(a) Central Diner and Grille, PA}
 \end{minipage}
 \quad
 \begin{minipage}[b]{0.45\linewidth}
    \includegraphics[scale=0.4]{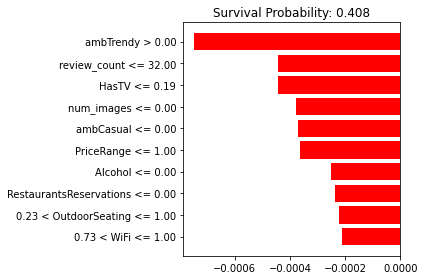}
   \caption*{(b) Planet Smoothie, AZ}
\end{minipage}
\caption{Top 10 most important business attributes for survival predictions of two restaurants. Important features for survival are coloured green and death are coloured red.}
\label{fig:lime_attr_exp}
\end{figure*}

%----------------------------------
% Feature Extraction
%----------------------------------
\subsection{Feature Extraction}
Unlike the features discussed earlier, these features are qualitative and describe several aspects of restaurants. 

\subsubsection{Pricing} pricing is an important aspect when deciding where to eat. The \textit{PriceRange}  attribute indicates the expensiveness of the restaurant's menu and is represented as an integer-based rating with 1 being the least expensive and 4 being the most expensive. 

\subsubsection{Ambience}
The ambience of a restaurant is measured as a boolean vector of nine dimensions including \textit{Romantic}, \textit{Intimate}, \textit{Classy},  \textit{Hipster}, \textit{Divey}, \textit{Touristy}, \textit{Trendy},  \textit{Upscale} and \textit{Casual}.

\subsubsection{Restrictions} Restrictions placed by the restaurants are also provided including \textit{DietaryRestrictions}, \textit{AlcoholServed}, \textit{GoodForKids}, \textit{DogsAllowed} and \textit{RestaurantAttire}.  

\subsubsection{Amenities} Several amenities available within the restaurant are also considered including \textit{OutdoorSeating}, \textit{BikeParking}, \textit{BusinessParking}, \textit{Wifi} and \textit{HasTV}.  

\subsubsection{Service} Restaurants also provide various auxillary services including \textit{TakesReservations}, \textit{HappyHour}. 

\subsubsection{Image and Review counts} For each restaurant, we also aggregate the total number of images and reviews it received from its customers. 

%----------------------------------
% Prediction Models
%----------------------------------
\subsection{Prediction Models}
\begin{figure}
    \centering
    \includegraphics[scale=0.5]{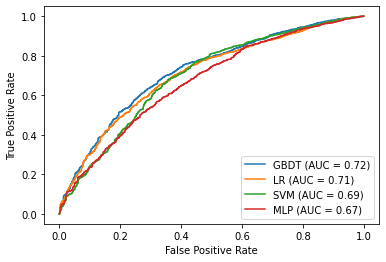}
    \caption{ROC curves for business attribute predictors. Positive class refers to survived label.}
    \label{fig:attr_roc_curve}
\end{figure}

Similar to section~\ref{sec:Geo_and_User}, we evaluate the survival predictions from these business attributes across four models including GBDT, MLP, LR and SVM. The resulting ROC curves are reported in Figure~\ref{fig:attr_roc_curve}. The GBDT model performs the best achieving an AUC score of 0.72. It is also interesting to note that, all models outperform their counterparts from figure~\ref{fig:roc_curve}. This suggests that business attributes are significantly better features at predicting business survival.

%----------------------------------
% Results
%----------------------------------
\subsection{Explaining Results}
We now refer back to the case of \textit{Central Diner} and \textit{Planet Smoothie}. As shown in Figure~\ref{fig:lime_attr_exp}, the former has been predicted with very high confidence of 98.8\% to survive mainly due to its trendy as well as classy ambience. In addition, it also has more than 149 reviews indicating that customers know in advance what they should be expecting. However, as seen in Figure~\ref{fig:lime_attr_exp}, the latter has been predicted to not survive with a confidence of 59.2\% mainly due to its trendy ambience and having less than 33 reviews. It can be observed that both restaurants had a trendy ambience, but one survives and the other does not. 
% TODO: Explain why this is
This can be attributed to fact that a trendy ambience is sought after by customers \textit{Central Diner}'s locality while the same cannot be deduced about \textit{Planet Smoothie}'s locality.
In contrast, there is a large disparity in the number of reviews that each of the respective restaurants had and this attribute is present in the top 3 attributes for the highest predictive power of survival prediction. Thus, the low number of reviews could account for the restaurant not being able to survive even though it has a trendy ambience.

%%%%%%%%%%%%%%%%%%%%%%%%%%%%%%%%%%%
% Linguistic modelling
%%%%%%%%%%%%%%%%%%%%%%%%%%%%%%%%%%%
\section{Linguistic Modelling}\label{sec:Linguistic}
The reviews posted by customers on Yelp provide detail of their experiences of the restaurants. This section further delves into extracting qualitative features from the text that can reveal a great deal of information regarding the survival of the business. Following the work of Lian et al. \cite{hetero_info}, two models are explored for survival prediction. As a further improvement, we implement a sentiment analysis model as we believe that this task better captures the impressions of customers from text.

%----------------------------------
% Feature Extraction
%----------------------------------
\subsection{Feature Extraction}
This section details various textual representations employed alongside the three models explored. 

\subsubsection{Review Text}
Each restaurant received at least three reviews in the Yelp dataset in which customers detail their visits. These review texts are shifted to lower case and pre-processed to remove stop words and punctuations. 

\subsubsection{Rating}
Along with the review, the customers must also provide a rating that is an integer between 1 to 5. To perform sentiment analysis, we follow
% TODO: Explain why 5 was left out. 
Zhang et al. \cite{char_level_cnn}
and transform the review ratings into polarity scores by considering ratings 1 and 2 as negative and 3 and 4 as positive. 

\subsubsection{Bag-of-Words}
The Bag-of-Words (BOW) feature involves determining the word counts across all the processed reviews for each restaurant. The recommended vocabulary size of 1000 words is utilised \cite{hetero_info}.

%----------------------------------
% Prediction Models
%----------------------------------
\subsection{Prediction Models}
Three prediction models are implemented across two prediction tasks namely survival prediction and sentiment analysis. 

% LimeTextExplainer
\begin{figure*}[ht!]
 \centering
 \begin{minipage}[b]{0.45\linewidth}
 \centering
 \includegraphics[scale=0.21]{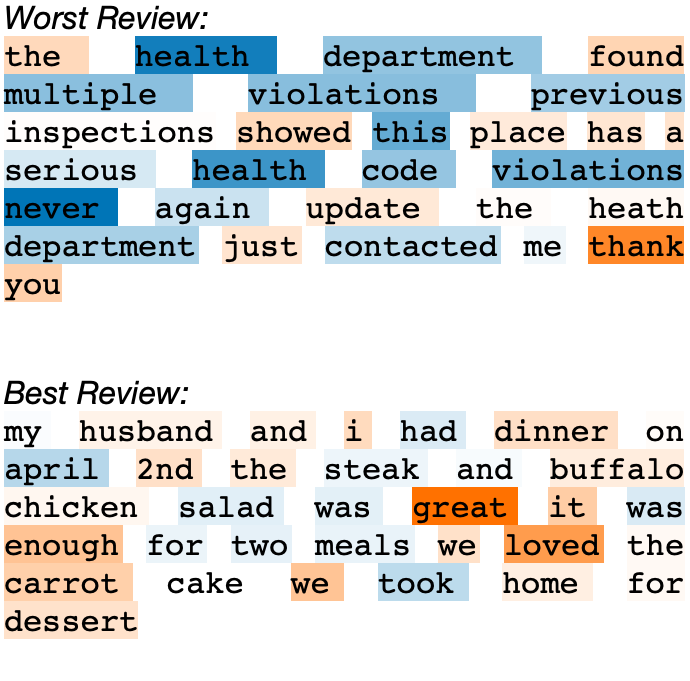}
  \caption*{(a) Central Diner and Grille, PA}
 \end{minipage}
 \quad
 \begin{minipage}[b]{0.45\linewidth}
 \centering
    \includegraphics[scale=0.21]{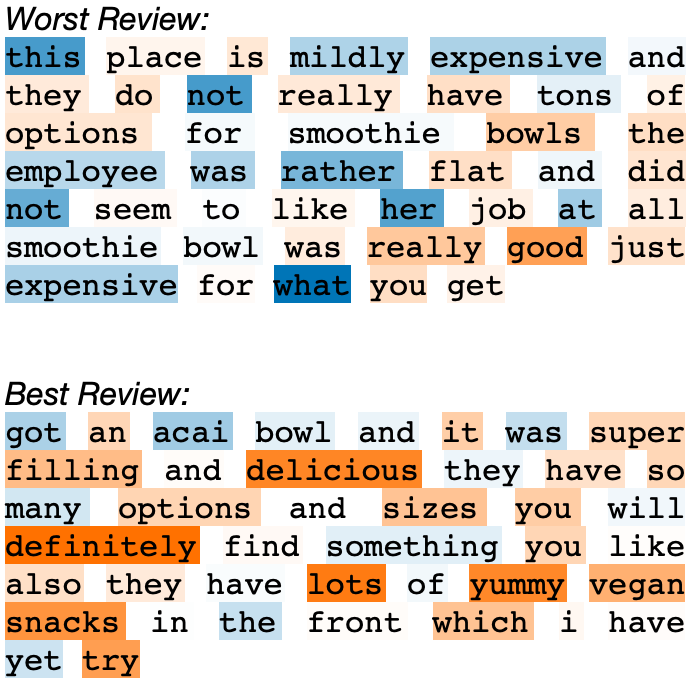}
   \caption*{(b) Planet Smoothie, AZ}
\end{minipage}
\caption{Sentiment analysis of the best and worst review for two restaurants. Important words for predicting survival are highlighted in orange and death in blue.}
\label{fig:lime_text_exp}
\end{figure*}

\subsubsection{Survival Prediction}
Following Lian et al. \cite{hetero_info}, we employ the BOW features of a restaurant to predict its survival. Since the prediction model employed is not mentioned, we examine GBDT and MLP. Lian et al. had also implemented a recurrent neural network (RNN) which did not perform as well as the BOW model. As future work, they also suggested the long short-term memory (LSTM) \cite{lstm} to capture long-term dependencies within the reviews~\cite{kwon_geolifecycle_imwut19}. Hence, we implemented a single layer LSTM model with learnable word embeddings. 

The LSTM model differs from the BOW model in terms of input features as the former directly takes the review texts. It may seem intuitive to utilise the most recent review as input. However, we believe that this might narrow the textual information. Contrarily, concatenating all reviews received by the restaurant may yield very large texts. Instead, we utilise the rating that accompanies each review in the dataset to determine the worst and the best review received for each restaurant. If there are multiple best/worst reviews with the same rating, we pick one randomly. The two reviews are then concatenated using \texttt{<SEP>} token. This provides the model with a more balanced set of opinions when predicting survival.

\subsubsection{Sentiment Analysis}
The sentiment analysis task is performed to demonstrate that the customers' perceptions are better captured by this task compared to applying survival prediction task. To this end, the same features and prediction models are employed for the sentiment analysis task with two differences. Firstly, the two prediction classes are now the polarity scores (positive or negative) of the reviews instead of the business survival. Since the polarity scores for the best and worst review for a restaurant are now different in this task, the two reviews are input to the LSTM model separately without the \texttt{<SEP>} token. 

% AUC ROC Survival
\begin{figure}[t!]
    \centering
    \includegraphics[scale=0.5]{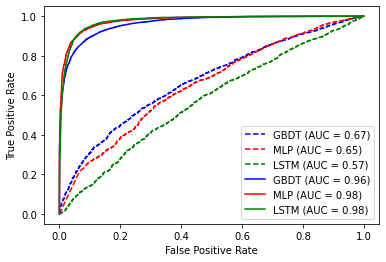}
    \caption{ROC curves for different linguistic models. Dashed lines refer to survival prediction models and solid lines refer to sentiment analysis models.}
    \label{fig:bow_lstm_surv}
\end{figure}

As shown in Figure~\ref{fig:bow_lstm_surv}, An immediate distinction in the performance of the models between the two tasks can be discerned. Applying the linguistic features for the sentiment analysis task significantly outperforms the survival prediction task. Additionally, LSTM performs the worst among the survival prediction models with an AUC score of 0.57, while it performs the best in sentiment analysis models with an AUC score of 0.98. Still, the GBDT and MLP models perform well on the survival prediction task with AUC scores of 0.67 and 0.65 respectively with a significantly smaller parameter size compared to LSTM. 

\vspace*{-0.1cm}

%----------------------------------
% Explaining Results
%----------------------------------
\subsection{Explaining Results}

This is further exemplified by the text explanations generated by \textit{LimeTextExplainer} for sentiment analysis from the LSTM model. Explanations were generated for the best and worst review of the two restaurants: \textit{Central Diner} and \textit{Planet Smoothie}. In both cases, the positive aspects of the two reviews laud their respective dishes: ``buffalo chicken salad'' at \textit{Central Diner} and ``acai bowl'' at \textit{Planet Smoothie}. Furthermore, the explanations also clearly highlight the avenues for the business owners to work on. For example, the business owners at \textit{Central Diner} must work toward mitigating the ``serious health code violations'' whereas the decision-makers at \textit{Planet Smoothie} will want to consider lowering their prices.

%%%%%%%%%%%%%%%%%%%%%%%%%%%%%%%%%%%
% Combined Models
%%%%%%%%%%%%%%%%%%%%%%%%%%%%%%%%%%%
\section{Combined Models}\label{Combined_Models}
The previous sections evaluated the performance of features from each feature set individually. In this section, we perform an ablation study through experimentation on different combinations of features and quantitatively evaluate the combination with the best predictive power. All the features are evaluated by applying them to the GBDT and MLP models. Majority voting is then performed for amongst upto four GBDT/MLP models for each set of features. Equal weights in voting are provided to each model from each feature set. 

\begin{table}[]
    \centering
    \begin{tabular}{lrr}
    \toprule
    Features &  GBDT &  MLP\\
    \midrule
    G    &  0.62 & 0.52 \\
    U    &  0.56 & 0.54 \\
    A    &  \textbf{0.72} & 0.67 \\
    L    &  0.65 & 0.67 \\
    GU   &  0.62 & 0.53 \\
    ALL &  0.51 & 0.54 \\
    -GU   &  0.61 & 0.64 \\
    -G  &  0.54 & 0.57 \\
    -U  &  0.51 & 0.55 \\
    -A  &  0.52 & 0.55 \\
    -L  &  0.61 & 0.60 \\
\bottomrule
\end{tabular}
    \caption{AUC performance of GBDT and MLP models using different feature combinations.}
    \label{tab:ablation_study}
\end{table}

The AUC scores of models utilising different combinations of features are reported in Table~\ref{tab:ablation_study}. Letters in the Feature column refer to individual features utilised - (G)eography, (U)ser Mobility, Business (A)ttributes and (L)inguistic. GU is a combination of geography and user mobility features. ALL refers to a model utilising all features. Feature names preceded by `-' refers to a model with features except for those mentioned e.g. -G model utilised user mobility, business attributes and linguistic features. The best-performing model is the GBDT model with business attribute features. In general, it is realised that combining features is not as informative to the prediction models. Within the combined models, the GBDT and MLP that employ the qualitative features of business attributes and BOWs perform better than models that employ the quantitative features. This suggests that the qualitative features are very crucial in predicting business survival.

%%%%%%%%%%%%%%%%%%%%%%%%%%%%%%%%%%%
% Conclusion and Future Work
%%%%%%%%%%%%%%%%%%%%%%%%%%%%%%%%%%%
\section{Conclusion and Future Work}\label{sec:Conclusion}

% Conclusion
In conclusion, this work has investigated developing interpretable classifiers to predict business survival by utilising four feature sets including geography, user mobility, business attributes and linguistics. Among them, empirical results demonstrate that the qualitative features including business attributes and language hold the highest predictive power for this task. The explanations generated by LIME with these features for two restaurants (\textit{Planet Smoothie} and \textit{Central Diner}) suggested highly intuitive reasons underlying the predictions made. 

Nonetheless, there are several shortcomings in this work. In the case of linguistic modelling, 
% aggregating the reviews of a restaurant to be input to the LSTM for survival prediction was found to be a challenging task and better approaches will be explored as future work. 
concatenation of the best and worst review received may not be ideal when there is a unbalanced number of positive and negative reviews. Instead of LSTM, Better prediction models could be employed that can take multiple texts as input.
Additionally, the combined models did not perform as well as expected which can be ascribed to the majority voting applied among the classifiers while future works can delve into more complex ensemble learning methods.

%%%%%%%%%%%%%%%%%%%%%%%%%%%%%%%%%%%
% Acknowledgement
%%%%%%%%%%%%%%%%%%%%%%%%%%%%%%%%%%%
\section*{Acknowledgments}\label{sec:Acknowledgement}
This research has been supported in part by the 5GEAR project (Decision No. 318927) and the FIT project (Decision No. 325570) from the Academy of Finland. 

% \clearpage

%%%%%%%%%%%%%%%%%%%%%%%%%%%%%%%%%%%
% References
%%%%%%%%%%%%%%%%%%%%%%%%%%%%%%%%%%%
\bibliographystyle{IEEEtran}
\bibliography{ref}

\end{document}